\documentclass{pasj00}
\color{black}
\begin{document}
\SetRunningHead{K. Sato et al.}{XMM-Newton Observation of IC~310}
\Received{2005 June 21}
\Accepted{2005 August 24}

\title{
XMM-Newton Observation of IC~310  in the Outer Region of the Perseus Cluster of Galaxies}

\author{
 Kosuke \textsc{Sato},\altaffilmark{1}
 Tae \textsc{Furusho},\altaffilmark{2} \\
 Noriko \textsc{Y.~Yamasaki},\altaffilmark{2}
 Manabu \textsc{Ishida},\altaffilmark{1} 
 Kyoko \textsc{Matsushita},\altaffilmark{3} \\
and Takaya \textsc{Ohashi}\altaffilmark{1}
}
 \altaffiltext{1}{Department of Physics, Tokyo Metropolitan University,\\
 1-1 Minami-Osawa, Hachioji, Tokyo 192-0397}
\email{ksato@phys.metro-u.ac.jp}
 \altaffiltext{2}{Institute of Space and Astronautical Science, Japan Aerospace Exploration Agency, \\
3-1-1 Yoshinodai, Sagamihara, Kanagawa 229-8510}
 \altaffiltext{3}{Department of Physics, Tokyo University of Science,\\
 1-3 Kagurazaka, Shinjuku-ku, Tokyo 162-8601}
\maketitle

\begin{abstract}

We present results from an XMM-Newton observation of the
head-tail radio galaxy IC~310 located in the southwest region of the
Perseus cluster.  The spectrum is well-fitted by an absorbed power-law
model with a photon index of $2.50 \pm 0.02$ with no significant
absorption excess.  The X-ray image shows a point-like emission at
IC~310 without any signs of a structure correlated with the radio halo
tail.  The temperature of the intracluster medium surrounding IC~310
declines as a function of distance from the cluster center, from $ kT
\sim 6$ keV in the northeast corner of the field of view 
to about 3 keV in the southwest region. 
Although we do not find any sharp edges in the surface brightness profile, 
a brightness excess over a  smooth $\beta$ model by about 20\% is seen. 
The temperature also rises by about 10\% in the same region. 
This indicates that the IC~310 region  is a subcluster probably 
infalling into the Perseus cluster, and the gas in front of IC~310 
towards the Perseus cluster is likely to be compressed 
by the large-scale motion, which supports the view that the  
IC~310 system is undergoing a merger.

\end{abstract} 

\section{Introduction}

Recent imaging and spectral observations of clusters of galaxies 
by Chandra and XMM-Newton have
significantly altered our view about clusters of galaxies. In
particular, dynamical effects of infalling subgroups have been
recognized in the form of cold fronts and shock fronts (e.g.,
\cite{Markevitch00}; Markevitch, Vikhlinin 2001; \cite{Vikhlinin01}). 
In the Coma cluster, for example, 
an X-ray image showed gas heating
between the NGC~4839 subcluster and the main cluster \citep{neumann01}. 
Such features clearly indicate that a subcluster merger is indeed
taking place, and we can study how the gas is heated and how the extra
energy is transferred into the surrounding space. High-sensitivity
imaging spectroscopy of subclusters in the cluster outskirts will, 
therefore, produce important knowledge about the evolution of clusters.

IC~310 ($z=0.0189$) is an S0 galaxy located in the southwest of the
Perseus cluster ($z=0.0183$) at an apparent  distance of about $\sim$1 
Mpc from the
cluster center. This galaxy is known as a head-tail radio galaxy 
(Sijbring, de Bruyn 1998), indicating that the radio lobe with a length about
350 kpc ($15'$) is extended to the southwest direction, parallel to the
line connecting IC~310 and the center of the Perseus cluster.  This
suggests that the galaxy may be moving toward the cluster center.  
de Bruyn and Brentjens (2005) have also detected significant 
polarized emission from 
a region close to IC~310 from low-frequency observations. 
\citet{schwarz92} suggest a substructure around IC~310 from
the PSPC image.  An X-ray observation of IC~310 region is, therefore,
an interesting case for studying any effect of subcluster motion
on the ambient plasma around IC~310. The X-ray halo, if it is
confirmed, can be used to probe the interacting features between the
galaxy and the cluster gas.

The galaxy IC~310 also has its own scientific interest, such as time
variability and Fe line feature. The
luminosity measured by ROSAT ($7.6\times 10^{42}$ erg s$^{-1}$,
\cite{sambruna99}) and ASCA ($\sim10^{43}$ erg s$^{-1}$,
\cite{furusho01a}) suggests that this galaxy holds an active nucleus.  Each
spectrum was fitted well by a single power-law model with a spectral
photon index of 3.7 and 2.1 with ROSAT and ASCA,
respectively. The temperature map of the Perseus cluster obtained with
ASCA showed an extended cool emission with $ kT \sim 4$ keV around
IC~310 \citep{furusho01b}.

This paper reports on the results from an XMM-Newton
observation of the IC~310 region. We use $H_0=70$ km s$^{-1}$ Mpc$^{-1}$, 
$\Omega_0$=0.3, $\Omega_{\Lambda}$=0.7 in this paper. 
At a redshift of 0.0189, 1$'$ corresponds to 23.7 kpc.
 
\section{Observation and Data Reduction}\label{sec:obs}

The XMM-Newton observation of IC~310 was carried out on
2003 February 26, for 29.4 ks. All of the EPIC cameras were operated in
the full frame mode, and the medium filter for MOS and the thin filter
for pn were used.  Data reduction and analysis were performed by
employing SAS version 6.0, CIAO 3.2, and HEAsoft version 5.3.

Data of this observation obtained with XMM-Newton was affected
by high-background flares.  The data-cleaning method used in
\citet{reiprich04} and Pratt and Arnaud (2002) were employed in 
event selection.  First, we checked the light curves in the high-energy 
bands of 10--12 keV for MOS and 12--14 keV for pn, where the
events were dominated by particle background due to soft protons.
Good time intervals (GTIs) without background flares were selected
from the light curves using conservative ``generic cuts'', in which
the count rate, $C$, was limited by $2.5(17.5) < C < 6.5(29.5)$ counts
s$^{-1}$ for MOS(pn), respectively.  Second, we applied the GTIs and
selected the photon events with patterns 0--12 for MOS and 0--4 for pn,
and flag = 0.  We calculated the count-rate distribution with 100~s
intervals, and then obtained new GTIs by requiring all of the count rates
to be within $\pm 3\sigma$ around the mean.  In this way, we iterated
the process until the variation of the number of usable data in a step
reached less than 5\% compared with the previous value.  The final
count rates were 3.26--5.31 counts s$^{-1}$ for MOS1, 3.10--5.04
counts s$^{-1}$ for MOS2, and 19.3--27.9 counts s$^{-1}$ for pn.
After the above data selection, the remaining exposure times were 25.0,
24.9, and 19.2 ks for MOS1, MOS2, and pn, respectively.

We further needed to subtract the non X-ray background and the cosmic
X-ray background (CXB).  In a spectral analysis of IC~310, we used
its outer annulus as the background, as described later.  For
analyzing the ICM, we used a background event data set created from
blank-sky observations in Read and Ponman (2003).  For the background files
we applied the same selection criteria for the source event.  The
source-to-background count rate ratio was 0.90 for MOS1, 0.94 for MOS2,
and 1.16 for pn calculated from the count rates in the 10--12 and
12--14 keV for MOS and pn. We rescaled the background spectra by
rewriting the BACKSCAL header keyword according to these ratios
\citep{fujita04}.  Noting that the pn camera is considerably sensitive
to background flares \citep{katayama04}, we therefore examined the
results by changing the background normalizations by $\pm 10\%$ to
assess any systematic uncertainties.

To make a correction for vignetting effects, the photon-weighting method of
\citet{arnaud01b}, as implemented in the SAS task ``evigweight'', was
applied to each source file and background file.  For each event, this
task computed the corresponding weight coefficient, defined as the
ratio of the effective area at the photon position and the energy to
the central effective area at that energy.  When extracting spectra
and images for specified energy bands, each event was weighted by this
coefficient.  Point sources were excluded by the SAS task ``edetect''
for all the event files. Throughout the paper we used the Galactic
value of $N_{\rm H}=1.33\times 10^{21}$ cm$^{-2}$ (Dickey, Lockman 1990) in
the direction of IC~310 as the absorption for spectral analysis.

\begin{figure}[!hb]
\centerline{\FigureFile(8cm,5cm){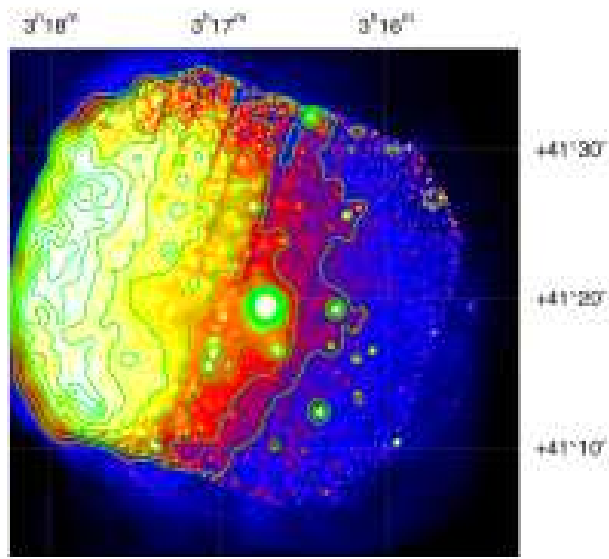}}
\caption{Combined MOS + pn image of ICM around IC~310 in the
0.5--10.0 keV energy range. X-ray images and contours 
show adaptively smoothed, 
background subtracted, 0.5--10.0 keV MOS+pn images.}
\label{fig:image-smth}
\end{figure}

An X-ray image of the whole MOS and pn fields in the energy band of
0.5--10 keV is shown in figure \ref{fig:image-smth}.  We made corrections for
exposure and vignetting and adaptively smoothed the image.  The gaps
between the CCD chips are still seen in the image.  The ICM emission
is obviously brighter towards the center of the Perseus cluster, which
is located in the northeast direction.  The bright source at the
center is IC~310, and apparently neither X-ray halo nor substructure is
associated with the galaxy in the image.

\section{Result}\label{sec:result}

\subsection{IC~310}

\begin{figure}[!hb]
\centerline{\FigureFile(8cm,5cm){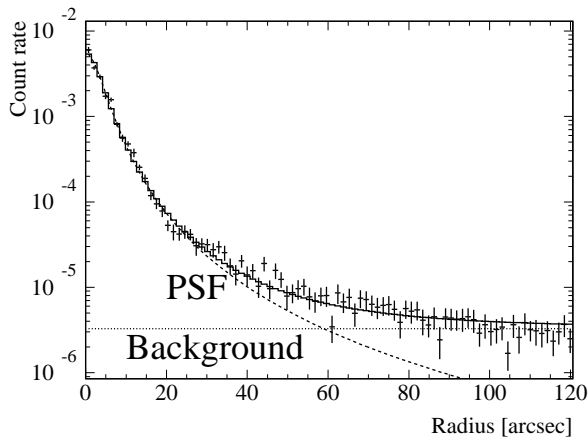}}
\caption{Radial profile of IC~310.
The MOS1 data were fitted with PSF and the background.
The dotted and dashed curves show the background level and 
the profile of the point spread function (PSF), respectively, 
which are best fit to  the radial profile. 
The solid curve is their sum.}
\label{fig:ic310-radial}
\end{figure}

We first looked into IC~310 and its immediate region
within a radius of $2'$.  In this region, since the brightness variation is
less than 5\%, we can assume the ICM distribution to be uniform. The
radial profile, centered on IC~310, for the MOS1 data of the 
exposure-corrected count rate in the energy band of 0.5--1 keV is shown in
figure \ref{fig:ic310-radial}.  The radial profile was fitted well
with the Point Spread Function (PSF) and a constant, which was the sum
of the NXB, CXB, and ICM emissions.  Thus, the X-ray emission is
considered to come from a point source, which is likely to be the
central AGN of IC~310.

\begin{figure}[!hb]
\centerline{\FigureFile(8cm,5cm){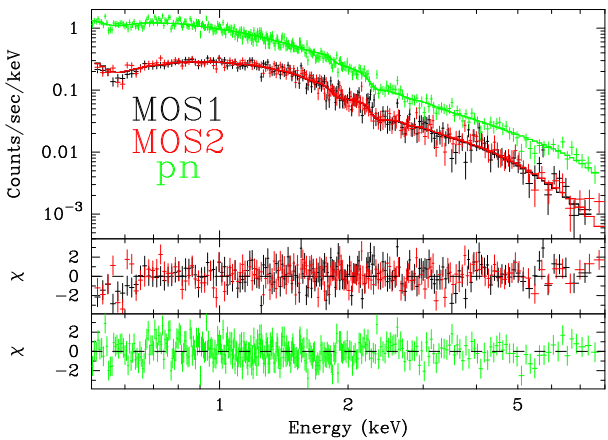}}
\caption{X-ray spectra of IC~310 from MOS1 (black), MOS2 (red), 
and pn (green).  
The data were fitted well with an absorbed power law with a photon  
index of 2.50. The fit was carried out simultaneously with three  
cameras.}
\label{fig:ic310-spec}
\end{figure}

We extracted the source spectrum from the central circle with a radius
of 2$^{\prime}$, and the background data were taken from an annulus
between radii of $2'-4'$. The resulting spectra for the different
instruments are shown in figure \ref{fig:ic310-spec}.  The spectra of
MOS1, 2, and pn were simultaneously fitted with an absorbed ($N_{\rm
H}=1.33\times 10^{21}$ cm$^{-2}$) power-law model. The best-fit photon
index is 2.50, and the normalizations are consistent within 9\%
between the detectors.  The obtained best-fit parameters are
summarized in table.\ref{tab:ic310}.  The flux and luminosity were
taken from the results of MOS1, whose values are consistent with the
past results obtained by ROSAT and ASCA
\citep{sambruna99,furusho01a}.  We derived an upper limit for the
equivalent width of the neutral Fe line (6.28 keV at $z= 0.0189$ ) 
to be $EW < 180$ eV at the 90\% confidence level.
\begin{table}[!ht] 
\begin{center} 
\caption{Best-fit parameters of the spectral analysis of IC~310.}
\begin{tabular}{cc} 
\hline \hline
Parameter & MOS+pn \\
\hline
$N_{\rm H}$ (~$10^{21}$ cm$^{-2}$~) & 1.33 (fixed) \\
photon index & 2.50$^{+0.02}_{-0.02}$  \\
reduced~$\chi^2$(degrees of freedom) & 1.12(688) \\
$F_{\rm X}$(0.5--2 keV) (erg cm$^{-2}$ s$^{-1}$) & $1.60~\times~10^{-12}~^{*}$ \\
$F_{\rm X}$(2--10 keV) (erg cm$^{-2}$ s$^{-1}$) & $1.38~\times~10^{-12}~^{*}$ \\
$L_{\rm X}$(0.5--2 keV) (erg s$^{-1}$)  & $2.05~\times~10^{42}~^{*}$\\
$L_{\rm X}$(2--10 keV) (erg s$^{-1}$)  & $1.14~\times~10^{42}~^{*}$\\
Fe line ($EW$) (eV) & $<$ 180~ \\
\hline 
\multicolumn{2}{l}{All of the errors are at the 90\% confidence level.}\\
\multicolumn{2}{l}{\footnotemark[*] The value obtained from MOS1; see text.}
\end{tabular}
\label{tab:ic310}
\end{center}
\end{table}
In order to examine the time variability of IC~310, we looked at the
light curve for the data in the central $2'$ (radius) circle, as
compared with those in the $2'-4'$ annulus for the background.  The
count rate integrated over 2 ks showed no significant variation
in excess of the statistical errors, which were about 5\% at the 90\%
confidence level.

\subsection{Intracluster Medium (ICM)}\label{sec:ICM}
\subsubsection{Surface brightness profile}
\begin{figure}[!hb]
\begin{center}
\FigureFile(7cm,5cm){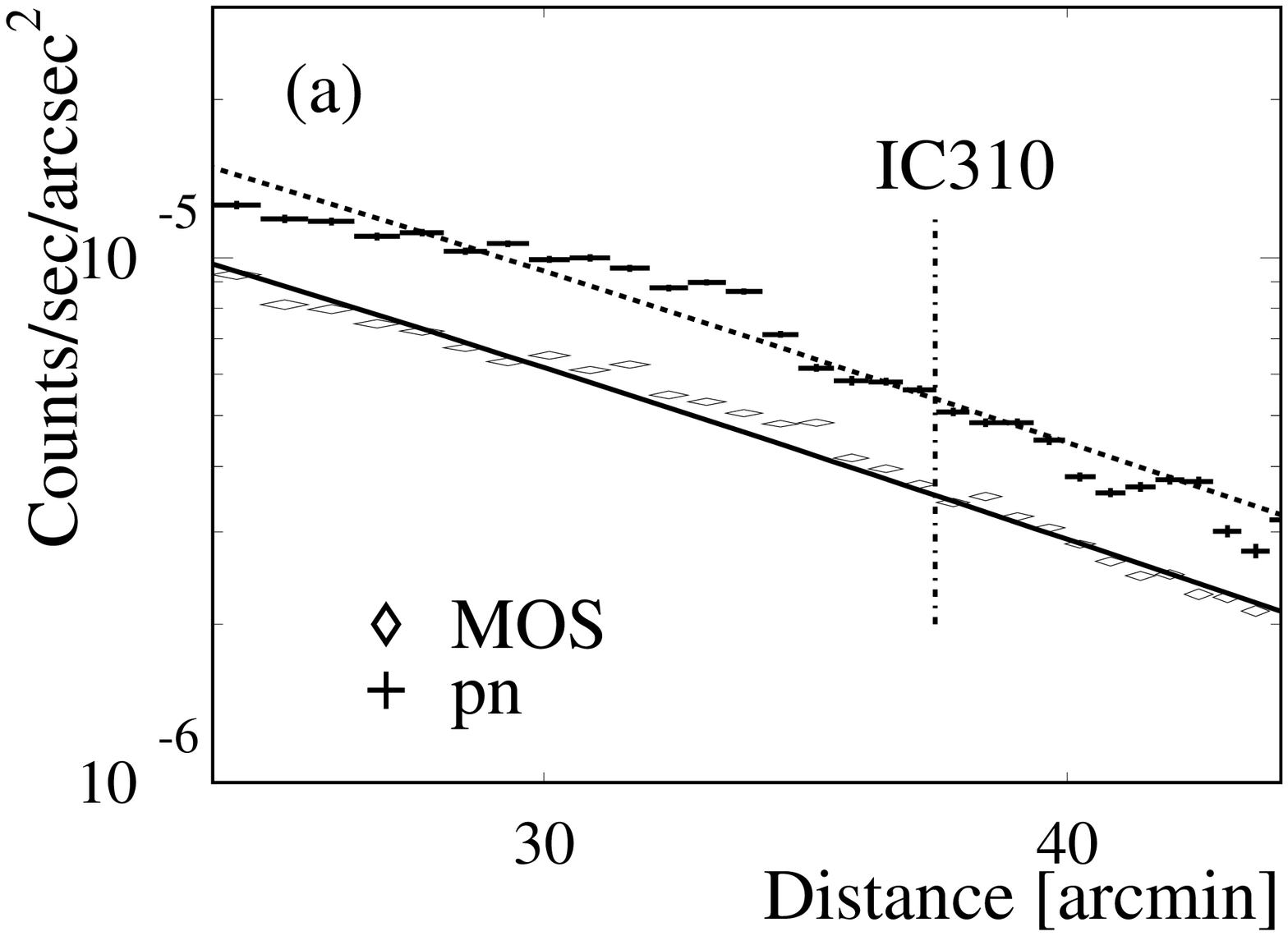}
\FigureFile(7cm,5cm){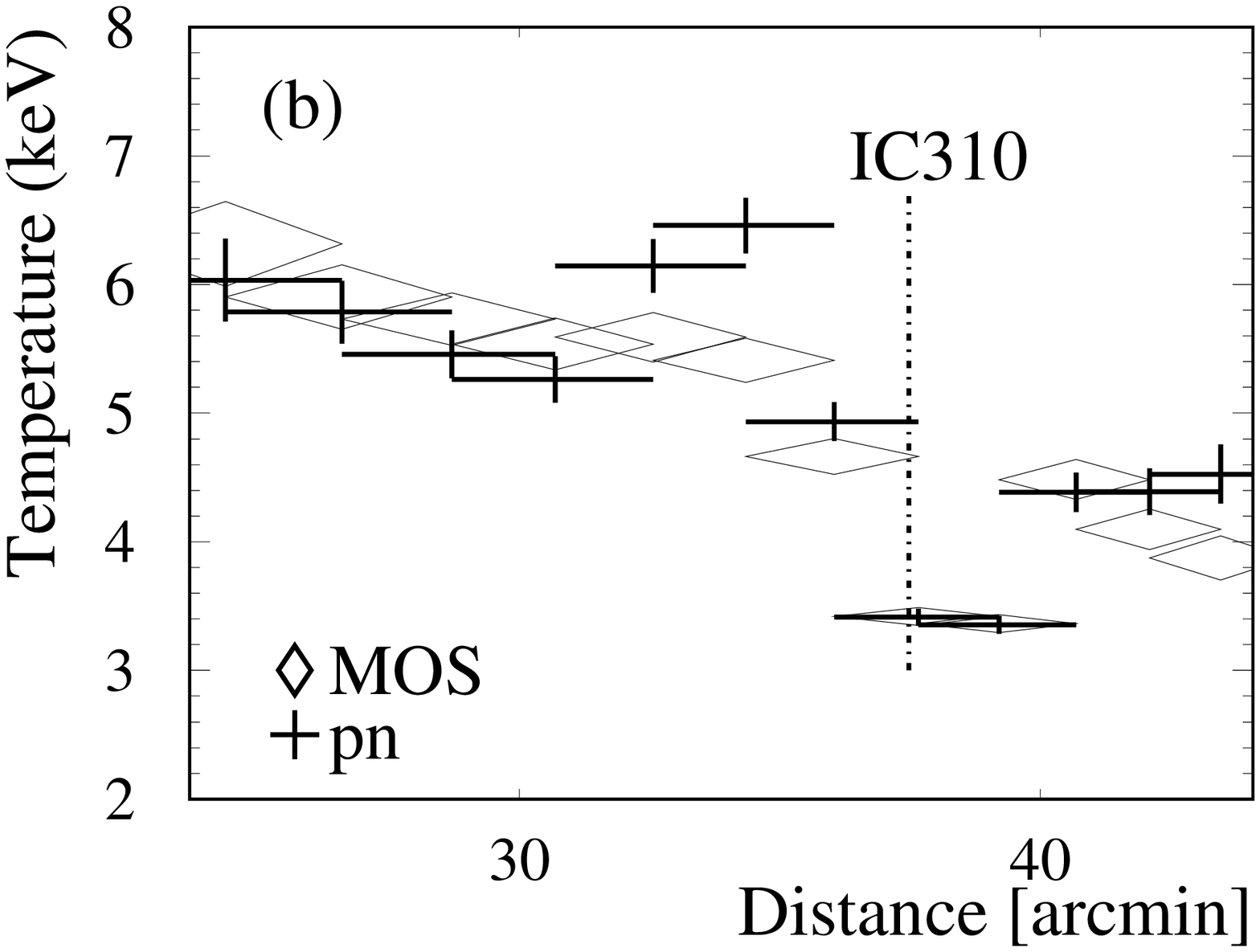}
\end{center}
\caption{(a) Surface brightness profile, plotted as a function of 
the distance from the center of the Perseus cluster, 
in the 0.5--10.0 keV band.  This
profile was background subtracted and corrected for vignetting.  The
solid and dotted lines in the left figure are the best-fit
$2\beta$-model for each camera.  (b) Temperature profile from the
hardness ratio between the energy ranges, 2--7.5 and 0.9--2 keV, for
MOS and pn for the ICM around IC~310.  }
\label{fig:sbandkt}
\end{figure}

The surface brightness profiles, plotted as a function of the distance
from the center of the Perseus cluster, after point-source
subtraction, are shown in figure \ref{fig:sbandkt} (a).  The energy
range is 0.5--10.0 keV\@. Events were binned in circular annuli
centered at the X-ray emission peak of the center of the Perseus
cluster.  Background subtraction was carried out separately for each
camera.  The solid and dotted lines in figure \ref{fig:sbandkt} (a)
indicate the best-fit 2-$\beta$ model for the total cluster emission
derived by \citet{furusho01a}. The values of $r_{\rm c}$ and $\beta$
were fixed at $r_{\rm c1} = 10'.1, r_{\rm c2} = 5'.0, \beta_1 = 0.64,$
and $\beta_2 = 1.16$. Only the normalization was a
free parameter.  The difference between the data and the 2-$\beta$ model
suggests that the observed data are flatter in the range $25'-30'$ and
a small excess is seen in the $30'-35'$ region.  
To look into the azimuthal dependence of the radial surface brightness 
profile, we divided the observed field into three evenly spaced sectors
centered at the Perseus cluster center with an opening angle 
of 16$^{\circ}$. The central line of the middle sector runs on IC~310. 
We compared  the radial brightness profiles of these sectors, but no
significant difference could be recognized in the intensity distribution for
the 3 sectors.
\subsubsection{Temperature profile from hardness ratio distribution}
\label{subsec:hr}
\begin{figure}[!hb]
\begin{center}
\FigureFile(7cm,5cm){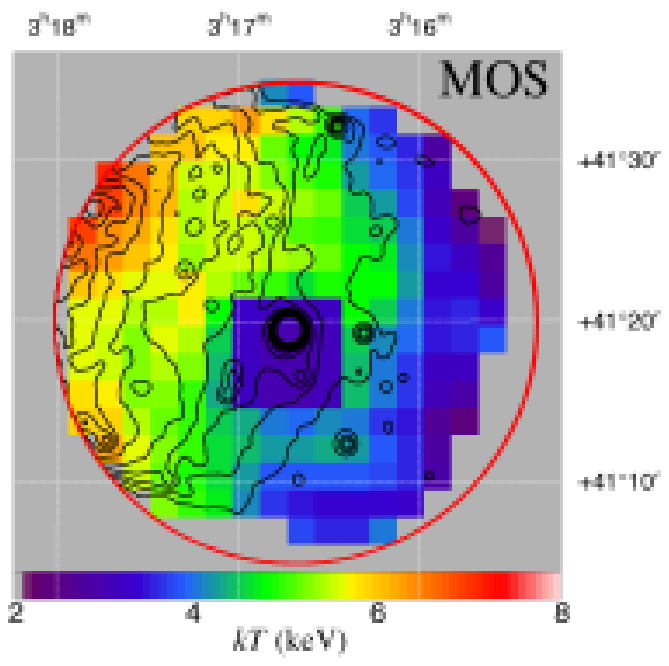}
\FigureFile(7cm,5cm){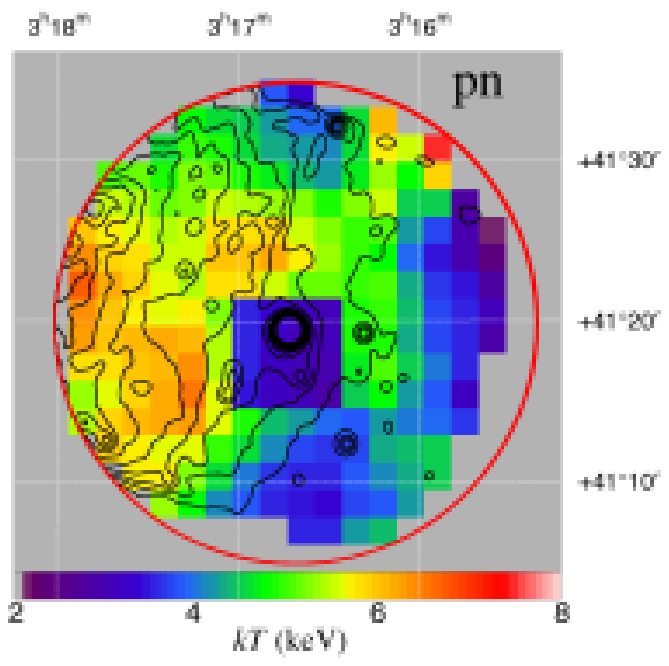}
\end{center}
\caption{Temperature maps obtained by MOS and pn 
overlaid with the same contour as in figure \ref{fig:image-smth}. 
The temperatures were calculated by HRs between below and above 2 keV.
The red circles indicate  the field of view of MOS.}
\label{fig:hardnessratio}
\end{figure}

We produced a temperature map of the projected ICM structure using
hardness ratios (HR).  After point source and background
subtractions, we produced images in two energy bands, 0.9--2.0 keV and
2.0--7.5 keV\@.  HR is defined as the ratio of the counts between
these bands.  Since the events were corrected
for the vignetting effect, the conversion factor from the HR to the
temperature is the same throughout the image.  We divided the region
into $1.7'\times1.7'$ cells for each image, took a running average
of $2\times2$ cells, and then calculated the HR values.  The temperatures
were derived by the response matrices at the center of the field of
view assuming an absorption of $N_{\rm H}=1.33\times 10^{21}$cm$^{-2}$
and a metal abundance of 0.3 solar.  The temperature maps for the combined
MOS data and the pn are shown in figure \ref{fig:hardnessratio}.  In both maps,
a temperature drop is clearly seen from $\sim$6 keV in the northeast
region to $\sim$3 keV in the southwest region. This feature is highly
significant because statistical errors on the temperature are less
than 10\% in the observed field. The effect of the background was 
examined by changing the background normalizations by $\pm10\%$ 
(see section \ref{sec:obs}). 
The obtained temperature values typically fluctuated by 10\%, 
but the relation that the temperature dropped by half 
remained almost unaffected.    
\subsubsection{Radial profile of temperature}
We also examined the radial profile of temperature along the direction
towards the center of the Perseus cluster.  The temperatures were
calculated from HRs in the same manner as described in
\ref{subsec:hr}, but without taking running averages.  The resulting
profile is shown in figure \ref{fig:sbandkt} (b).  The temperature
declines by 30\% from 6 keV at $r=25'$ to 4 keV at $r=45'$.  As can be seen
in the surface-brightness profile, a similar excess around $r=35'$ is
also noted in the temperature profile.  We also examined the possible
azimuthal dependence in the temperature profile in the same way as we
did for the surface-brightness profile, but no significant difference
was seen in the temperature profile over the  
statistical uncertainty, which is on the order of $\sim$10\%.
\subsubsection{Temperature and abundance profile from spectral analysis}
\begin{figure}[!hb]
\begin{center}
\FigureFile(7cm,5cm){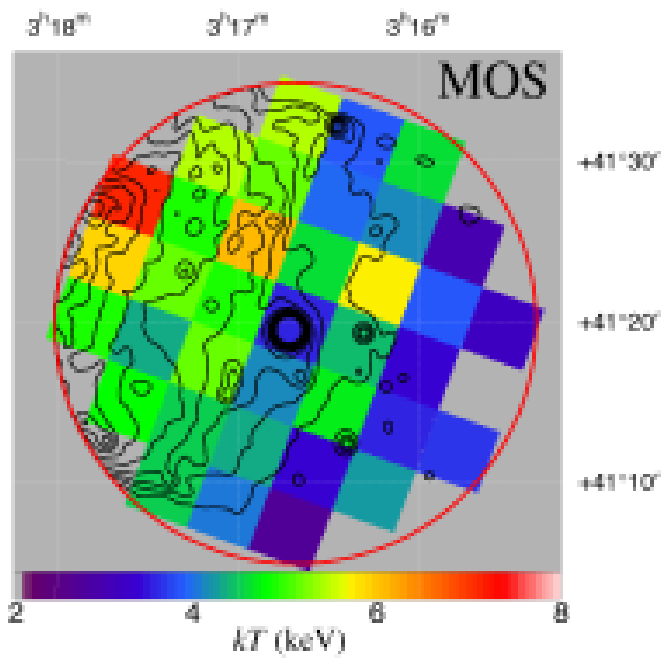}
\FigureFile(7cm,5cm){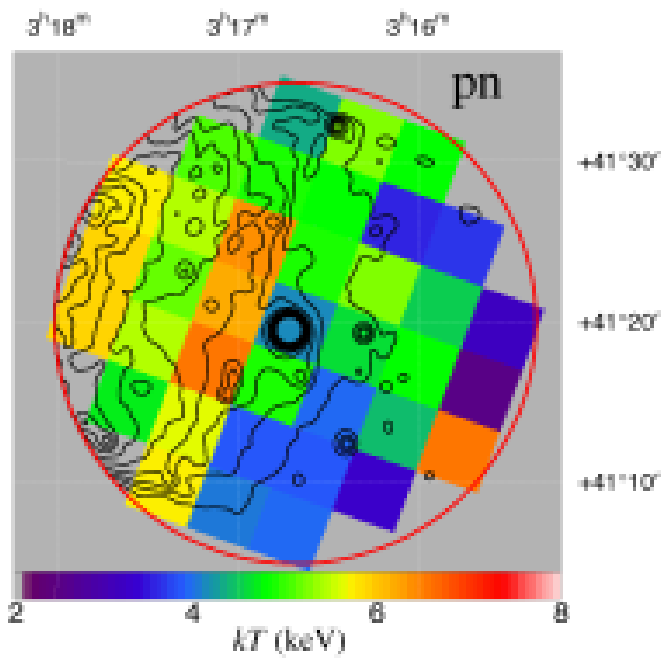}
\end{center}
\caption{Temperature map based on the spectral fit in the energy
range 0.9--7.5 keV\@.  The MOS and pn spectra were fitted with an
absorbed MEKAL model.  The red circle is the field of view of
MOS\@.}
\label{fig:temperature-fit}
\end{figure}

\begin{figure}[!hb]
\begin{center}
\FigureFile(7cm,5cm){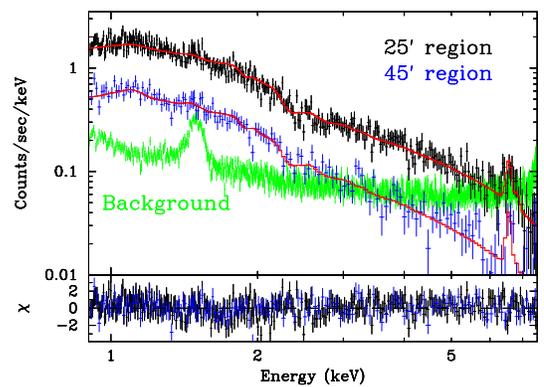}
\end{center}
\caption{X-ray spectrum of ICM around the $25'$ and the $45'$ region 
from the cluster center with pn camera. The data were fitted with 
the absorbed MEKAL model.}
\label{fig:spectrum-icm}
\end{figure}

\begin{figure}[!hb]
\centerline{\FigureFile(8cm,5cm){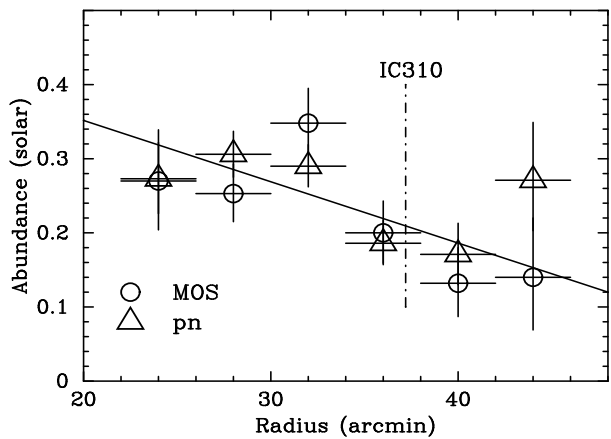}}
\caption{Radial abundance profile from a spectral fit 
in 0.9--7.5 keV~.
The MOS and pn spectra were fitted with an absorbed MEKAL model.
The solid line shows the best-fit linear model.}
\label{fig:abundance-fit}
\end{figure}

In order to obtain the temperature and metal abundance more accurately, we
carried out spectral fits for each $4'\times4'$ square region.  The
MOS and pn spectra were separately accumulated and fitted
simultaneously with an absorbed MEKAL model, both in the 0.9--7.5 keV
band in figure \ref{fig:temperature-fit}.  
The absorption was fixed at a Galactic value of $N_{\rm
H}=1.33\times 10^{21}$ cm$^{-2}$ in the direction of IC310.  
We show the resultant temperature map in figure \ref{fig:temperature-fit}, 
X-ray spectra with
pn camera for two annular regions in figure \ref{fig:spectrum-icm}, 
and the radial
abundance profile in figure \ref{fig:abundance-fit}, respectively.
The temperature map again indicates a drop from $\sim
6$ keV in the northeast region to $\sim 3$ keV in the southwest
region, as can be seen in the map calculated from the HR (figure
\ref{fig:hardnessratio}). The errors for the temperature are typically
20\% at the 90\% confidence level.  
When we varied the background normalization by $\pm10\%$, 
the temperature relation between the northeast and the southwest regions 
was unaffected, as can be seen in  section \ref{subsec:hr}.
The spectra in figure \ref{fig:spectrum-icm} shows the ICM
emission around the $25'$ and the $45'$ regions from the cluster
center, together with the non X-ray background, taken with the pn
camera. The ICM spectra are fitted with the absorbed MEKAL model, and
the temperatures are $5.7 \pm 0.3$ keV and $3.8\pm 0.3$ keV, for the inner
and outer regions, respectively. 
We also found that the metal
abundance declines from about 0.3 solar in the northeast region to
about 0.15 solar in the southwest region. This feature is also
statistically significant at the 90\% confidence level.  If these data
points are tested against a constant-value model, we obtain a reduced
$\chi^2 > 2$, whereas it reduces to about 1.3 for a linearly declining
model.

\section{Discussion}\label{sec:discuss}
\subsection{IC~310}
The X-ray radial profile of IC~310 was fitted well by the PSF of a
point source.  The energy spectrum was fitted with an absorbed
power-law model with a rather steep photon index of 2.5, and did not
exhibit Fe lines ($EW < 180$ eV)\@.  These features indicate that
the emission of IC~310 originates from the central AGN of the BL Lac-type 
object.  However, the time variability was not detected during our
observation, and the flux was consistent with the level measured during
the past 10 yr by ASCA and ROSAT.  We next estimated
the possible X-ray emission from hot interstellar medium in the
elliptical galaxy IC~310 based on the $L_{\rm X}-L_B$ relation
for elliptical galaxies \citep{canizares87,matsushita01,
yamasaki02}. The optical luminosity of IC~310 was shown to be 
$L_B \sim~1\times10^{10}L_{\odot}$. This corresponds to an X-ray halo
luminosity of $L_{\rm X} \sim~10^{39}$ erg s$^{-1}$, which is three
orders of magnitude lower than the that of the observed IC~310 value. This
indicates that the X-ray emission of IC~310 is dominated by the AGN
component, and the halo emission, if at all present, is masked by the
PSF of the strong point-source emission. With much higher angular or
spectral resolution, one may be able to detect the X-ray emission from a
hot-halo associated with this galaxy.

VLA observations indicate a radio lobe in IC~310, and the radio flux
densities at 49 cm $S_{49 \rm cm}$ are reported to be about 1.3
Jy (Sijbring, de Bruyn 1998).  We looked into the radio lobe region reported
in Sijbring and de Bruyn (1998), whose size is $15' \times 4'$, in the present
field, and searched for any flux excess by comparing the flux with
those in the immediate surrounding region.  We detected no significant
excess in the X-ray flux from the radio lobe; the upper limit is
shown in table.\ref{tab:radiolobe}.  If we assume that X-rays are
emitted through the inverse Compton process with 2.7 K photons by the
same relativistic electrons responsible for the radio lobe, the
strength of the magnetic field is constrained.  Following a
prescription given by Harris and Grindlay (1979), we derived the lower 
limit of the magnetic field strength to be $B>1 \mu$G\@.

\begin{table}[!ht] 
\begin{center} 
\caption{Upper (lower) limits of the excess of the radio lobe region 
at the 2$\sigma$ confidence level.}
\begin{tabular}{c}
\hline
\hline 
 $F_{\rm X}$(0.5--10 keV) [erg cm$^{-2}$ s$^{-1}$] \\
\hline
 $<~6~\times~10^{-14}$ \\
\hline
\hline
 $L_{\rm X}$(0.5--10 keV) [erg s$^{-1}$] \\
\hline 
$<~5~\times~10^{40}$ \\
\hline
\hline
 Magnetic filed [$\rm \mu$G] \\
\hline
  $>1$ \\
\hline 
\end{tabular}
\label{tab:radiolobe}
\end{center}
\end{table}

\subsection{ICM around IC~310}
The observed temperature in the present field is lower than the level
at the central region of the Perseus cluster, $\sim 8$ keV, excluding the
cool core.  The metal abundance is also lower than the central values
of $\sim 0.6$ solar \citep{ezawa01, churazov03}.  The ICM temperature
within the observed region also drops from $\sim6$ keV at the
northeast region to $\sim3$ keV at the southwest region.  These
significant variations in the temperature and metal abundance confirm the
results reported by \citet{furusho01b}, who revealed the temperature
gradient in the whole Perseus cluster based on ASCA
observations.  ASCA data indicated that the temperature in the
west region of the Perseus cluster dropped from $\sim8$ keV at $\sim
0.5$ Mpc from the center of the cluster to $\sim 5$ keV at $\sim 1$
Mpc.  Some clusters were recently investigated as far as $\sim 1$
Mpc from the center; we note that in some systems (e.g., A 1413,
A 478, A 2029, etc.), the temperature of ICM decreases by half from the
peak location (usually around $\sim0.3$ Mpc from the cluster center)
to the $\sim1$ Mpc region \citep{arnaud05,pointecouteau05,Vikhlinin05}.
Massive near-by clusters (e.g., the Coma cluster, the Virgo cluster)
also show a temperature drop by $\sim 40\%$ from the central region to
the $\sim0.5$ Mpc region \citep{kikuchi00,arnaud01a,shibata01}.  The
present case suggests that the infall of small and cool systems may be one
of the possible causes of creating a lower ICM temperature in the outer
region.

\citet{schwarz92} showed that the X-ray emission around IC~310 is more
extended than that of a point source, when a smooth $\beta$-model was fitted
and subtracted from a large-scale image, based on ROSAT
observations.  This implies that the IC~310 system is likely to be a
subcluster or a group of galaxies. ASCA observations reveal a
temperature drop at the region 
between the cluster center and IC~310 \citep{furusho01b}.
These results suggest that the ICM in this region is subject to heating due
to infall of the IC~310 system into the main Perseus cluster.

In the present observation, we did not detect any sharp structures, such as
cold fronts or strong shock fronts in the X-ray image.  We found,
however, an excess of surface brightness in the region between
IC~310 and the center of the Perseus cluster, as shown in figure
\ref{fig:sbandkt}.  A slight excess of the temperature is also seen in
the same region.  These features suggest that the ICM in this region
is compressed by the pressure of the infalling gas associated with the
IC~310 subcluster.  The pressure, $P$, and the emissivity, $I$, can be
expressed in terms of the electron density, $n_{e}$, 
and the temperature, $T$, as
$P\propto n_{\rm e}T$ and $I\propto n_{\rm e}^{2}\sqrt{T}$, which gives
$P\propto I^{1/2}T^{3/4}$.  The observed excess in the surface
brightness around IC~310 is $\sim 20\%$, compared with the smooth ${\rm
\beta}$ model profile for the whole cluster.  The temperature excess
is about 10\% compared with the smooth trend of the temperature
gradient in the present field of view.  Thus, the pressure in this
region is estimated to be $15\%$ higher than the surrounding region.

We can break down the general merging process into two steps: first,
the local temperature of the ICM is raised by shock heating or
compression heating, and then the temperature of the whole cluster
gradually increases due to growth of the gravitational potential
after a merger. The observed small hump in the brightness and
temperature around IC~310 suggests that the system is in a stage
between the first and second steps.  If the infall velocity of the
subcluster is close to the sound velocity in the region of the
pressure excess ($\sim 1200$ km s$^{-1}$), the crossing time for a
distance of 100 kpc ($4'$) is $\sim 1 \times 10^{8}$ yr.  In contrast,
the time scale for thermalization of the whole Perseus
cluster, estimated by the sound-crossing time over about 1 Mpc 
for a 5 keV plasma, is $\sim10^{9}$ yr.
   
\section{Conclusions}
We have presented the results from an XMM-Newton observation of
the radio galaxy IC~310 and the surrounding ICM\@.  The X-ray image of
IC~310 is point-like, and the spectrum was fitted by an absorbed
power-law model with a photon index of $2.50 \pm 0.02$ with no significant
absorption at the source.  An X-ray halo associated with this elliptical
galaxy was not detected.  No significant X-ray feature correlated with
the radio halo extending in the southwest direction was seen, and we
derived the lower limit for the magnetic field to be $>1\ \mu$G\@.

The emission from the ICM of the Perseus cluster filled the field of
view of EPIC detectors, and we detected a systematic change of the
temperature as a function of the distance from the center of the
Perseus cluster, from $ kT \sim 6$ keV in the northeast to $\sim 3$
keV in the southwest within the field of view.  No sharp edges were
seen in the brightness and temperature maps, but we recognized excess
features in their radial profiles in the region between the center of
the Perseus cluster and IC~310. In this region, the ICM is likely to be
compressed by the infalling subcluster.  The metal abundance in the
ICM also decreases gradually from about 0.3 solar in the northeast
region to about 0.1 solar in the southwest region.
\\

The authors thank the referee for providing valuable comments.
This research  was supported in part by Grants-in-Aid by 
the Ministry of Education, Culture, Sports, Science 
and Technology (14079103, 15340088).

\end{document}